\documentclass[letter]{aa}  

\usepackage{graphicx}
\usepackage{txfonts}
\newcommand{\um}{$\mu$m} 

\newcommand{\lya}{Ly$\alpha$}
\newcommand{\ha}{H$\alpha$}
\newcommand{\hb}{H$\beta$}
\newcommand{\oiii}{[O\,{\sc iii}]}
\newcommand{\oiiio}[1]{[O\,{\sc iii}]$\lambda\lambda${#1}} 
\newcommand{\mgii}{Mg\,{\sc ii}}
\newcommand{\civ}{C\,{\sc iv}}

\newcommand{\nii}{[N\,{\sc ii}]} 
\newcommand{\niio}[1]{[N\,{\sc ii}]$\lambda\lambda${#1}}

\newcommand{\oii}{[O\,{\sc ii}]} 
 
\newcommand{\sii}{[S\,{\sc ii}]} 
\newcommand{\siio}[1]{[S\,{\sc ii}]$\lambda\lambda${#1}} 

\newcommand{\Nh}{$N_\mathrm{H}$}
\newcommand{\edens}{$n_e$}
\newcommand{\Te}{$T_e$}
\newcommand{\zsun}{Z$_\odot$}
\newcommand{\msun}{M$_\odot$}

\begin{document}

   \title{Direct high-resolution observation of feedback and chemical enrichment in the circumgalactic medium at redshift $z \sim 2.8$}

   \titlerunning{SMM J02399-0136 CGM [O III]}

   \author{
         Bo Peng \inst{1} 
         \and
         Fabrizio Arrigoni Battaia \inst{1} 
         \and
         Amit Vishwas \inst{2} 
         \and
         Mingyu Li \inst{3} 
         \and
         Edoardo Iani \inst{4} 
         \and
         Fengwu Sun \inst{5} 
         \and
         Qiong Li \inst{6} 
         \and
         Carl Ferkinhoff \inst{7} 
         \and
         Gordon Stacey \inst{2,8} 
         \and
         Zheng Cai \inst{3} 
         \and
         Rob Ivison \inst{9,10} 
         }

   \institute{
         Max Planck Institute for Astrophysics,
         Karl-Schwarzschild-Straße 1, Garching bei München, 85748 BY, Germany
         \email{bopeng@mpa-garching.mpg.de}
         \and
         Cornell Center for Astrophysics and Planetary Science, Cornell University,
         122 Sciences Drive, Ithaca, 14850 NY, United States
         \and
         Department of Astronomy, Tsinghua University, 
         Beijing, 100084 People's Republic of China
         \and
         Kapteyn Astronomical Institute, University of Groningen, 
         P.O. Box 800, Groningen, 9700 AV Netherlands
         \and
         Center for Astrophysics, Harvard \& Smithsonian, 
         60 Garden Street, Cambridge, 02138 MA, United States
         \and
         Jodrell Bank Centre for Astrophysics, University of Manchester,
         Alan Turing Building, Oxford Road, Manchester, M13 9PL United Kingdom
         \and
         Department of Physics, Winona State University, 
         175 West Mark Street, Winona, 55987 MN, United States
         \and
         Department of Astronomy, Cornell University, 
         122 Sciences Drive, Ithaca, 14850 NY, United States
         \and
         European Southern Observatory, 
         Karl-Schwarzschild-Straße 2, Garching bei München, 85748 BY, Germany
         \and 
         Institute for Astronomy, University of Edinburgh, Royal Observatory, 
         Blackford Hill, Edinburgh, EH9~3HJ, United Kingdom
         }

   \date{Received 14 October 2024; accepted 4 January 2025}

  \abstract
   {
   The circumgalactic medium (CGM) plays a vital role in galaxy evolution, however, studying the emission from CGM is challenging due to its low surface brightness and the complexities involved in interpreting resonant lines such as Lyman-alpha (\lya{}). 
   }
   {
   The near-infrared coverage, unprecedented sensitivity, and high spatial resolution of the \textit{James Webb} Space Telescope (JWST) enable us to study the optical strong lines associated with the extended \lya{} ``nebulae'' at redshifts of 2--3. 
   These lines serve as diagnostic tools to infer the physical conditions in the massive CGM gas reservoir of these systems. 
   }
   {
   In deep medium-band images taken by the JWST, we serendipitously discovered the \oiii{} emission from the CGM surrounding a massive interacting galaxy system at a redshift of $z \sim 2.8$, known to be embedded in a bright extended (100 kpc) \lya{} ``nebula.'' 
   }
   {
   This is the first time that the \oiii{} lines have been detected from a \lya{} ``nebula.'' The JWST images reveal that the CGM gas actually resides in narrow ($\sim$ 2.5 kpc) filamentary structures with strong \oiii{} emission, tracing the same extent as the \lya{} emission. 
   An analysis of the \oiii{} suggests that the emitting CGM is fully ionized and is energetically dominated by mechanical heating. 
   We also find that the inferred density and pressure are higher than those commonly predicted by simulations of the CGM. 
   }
   {We conclude that the observed CGM emission originates from the gas expelled by the episodic feedback processes, cooling down and enriching the CGM, while traveling a distance of at least 60 kpc. 
   These observations demonstrate how intensive feedback processes shape gas distribution and properties in the CGM around massive halos. 
   While access to such deep, high-resolution imaging opens up a new discovery space for investigating the CGM, it also challenges numerical simulations with respect to explaining and reproducing the exquisitely complex structures revealed by the observations. }

   \keywords{ISM: jets and outflows -- 
                Galaxies: intergalactic medium --
                Galaxies: active -- 
                Galaxies: interactions
               }

   \maketitle
%

\section{Introduction}
\label{sec:intro}

The circumgalactic medium (CGM) fuels the growth of host galaxies and recycles the material ejected by feedback resulting from star formation or active galactic nucleus (AGN) activity. 
The physical conditions and chemical enrichment of the CGM therefore reflect the evolutionary history of the galaxy and herald its further growth \citep{oppenheimer10,tumlinson17}. 
Using the \lya{} line, past studies have had great success in finding the ``cool'' ($\sim 10^4$ K) phase of the CGM \citep[e.g.,][]{cantalupo14,martin15,borisova16,wisotzki16}, which is hypothesized to be the gas being accreted onto the host galaxies \citep{keres05,sancisi08}. 
The extended \lya{} emission is now routinely found around galaxies at a redshift of $z > 2$, where the \lya{} line is accessible from the ground, and can extend to beyond 100 kpc (\lya{} ``nebulae'') around quasars \citep{borisova16,battaia19,cai19,farina19,li24} or galaxy (proto)clusters \citep{matsuda11,umehata19,daddi22}. 

However, the interpretation of \lya{} is complicated by its resonant nature and efforts have been invested in exploring other lines, including UV resonant lines, such as \civ{} and \mgii{}, as well as optical strong lines including \ha{}, \oii{}, and \oiii{}, to study the ``cool'' CGM gas \citep[e.g.,][]{rupke19,helton21,li21,zhang23,langen23,johnson24}. 
With a clear dependence of the line intensity on the collisional rate between gas particles, the optical strong lines can be used to estimate the physical condition of the CGM gas. 

Among these optical lines, \oiiio{4959,5007\AA} are frequently used in identifying the spectral signatures of AGN outflows \citep{zakamska14,zakamska16,vietri18,perrotta19}. 
However, direct observations of spatially extended \oiii{} emission are sparse, mainly at low redshifts \citep{greene12,liu13,husemann14,fischer18,helton21,nielsen24}. 
At high redshifts, the \oiii{} outflows are occasionally seen on spatial scales up to $\sim$ 20 kpc \citep[e.g.,][]{nesvadba08,alexander10,cano12,harrison12,carniani15,wylezalek22,solimano24}. 
However, the sizes of these \oiii{} outflows are still much smaller than that of the cool CGM gas reservoir revealed by \lya{}, due to the limited observational capability in terms of near-infrared coverage, surface-brightness sensitivity, spectral resolution, and field of view. 
However, such observations can now be achieved with JWST narrow- or medium-band imaging, filling the gap between the \lya{} ``nebulae'' and the telltale optical strong lines. 

\begin{figure*}
\centering
\includegraphics[width=\textwidth]{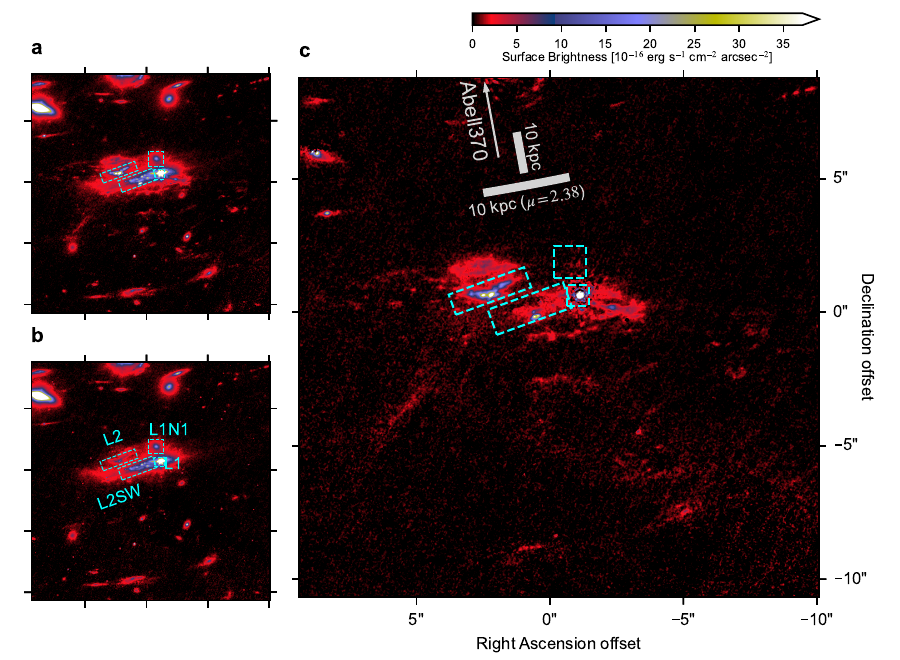}
\caption{20''$\times$20'' postage stamps of the SMM J02399-0136 system {\textbf a}, the JWST/NIRCam F182M image covering the \oiii{} emission. {\textbf b}, the interpolated stellar continuum based on the adjacent F150W and F210M images. {\textbf c}, the \oiii{} line emission map produced by subtracting {\textbf b} from {\textbf a}. Various continuum sources of the system identified in the previous studies are marked by the dashed cyan boxes. The grey arrow points towards the center of the gravitational lensing potential of Abell 370 \citep{richard10,gledhill24}, and the two grey scale bars show 10 kpc in the physical scale orthogonal to and along the lensing shear.}\label{fig:oiii_map}
\end{figure*}

The target of this paper is the galaxy group SMM J02399-0136 (R.A.=02h39m51.9s, Dec=-01d35m59s) at redshift $z$=2.808. 
It is lensed by the galaxy cluster Abell 370 ($z$=0.375) with a magnification factor of $\mu$ = 2.38 \citep{ivison10}.
The gravitational lensing stretches the shape of the observed image along the shear direction (labeled in Fig.~\ref{fig:oiii_map}), while preserving the shape in the orthogonal direction as well as the intrinsic surface brightness (Appendix~\ref{sec:distribution}). 
The group consists of a broad absorption line quasar (L1), a dusty star-forming galaxy (DSFG; L2SW), an irregular companion galaxy (L2), and a projected companion (L1N1) \citep{vernet01,ivison10,aguirre13}. 
Originally discovered as the first sub-millimeter selected galaxy \citep{ivison98}, this system is forming stars at a rate of at least 400 \msun{} yr$^{-1}$, taking place mostly in the DSFG \citep{ferkinhoff15}. 
The kinematics traced by molecular gas \citep{frayer18} and the shape of the stellar continuum suggest the DSFG has a rotating disk viewed edge-on, and the extended diffuse stellar light to the east is likely the tidal remnant of the ongoing merging between the DSFG and the quasar. 
This system also hosts one of the first \lya{} ``nebulae'' discovered through long-slit spectroscopy \citep{ivison98}, which was later found to extend to a scale of 100 kpc and host a large amount of ``cool'' CGM gas in integral field unit (IFU) observations \citep{li19}.
In addition, evidence of extended ionized and molecular gas was found in multi-wavelength observations \citep{genzel03,ivison10,ferkinhoff15,frayer18,vidal21}. 

In this Letter, we report a serendipitous discovery of \oiii{} emission in the CGM of SMM J02399. 
The observational data of \oiii{} and \lya{} are described in Sect.~\ref{sec:data}. 
Section~\ref{sec:results} describes the main discovery in the JWST data and its comparison to the \lya{} map. 
The implication of the discovery, including the physical condition of the emitting gas and source of \lya{} photon in the \lya{} nebula, as well as the processes that may be responsible for the morphology and power of the CGM emission, are discussed in Sect.~\ref{sec:discussion}. 
Detailed calculations and elaborate descriptions are given in the appendix, while our general conclusions are discussed in Sect.~\ref{sec:discussion}. 
We conclude the paper with a brief summary in Sect.~\ref{sec:summary}. 

\section{Observational data}
\label{sec:data}

This work utilizes the medium-band images obtained in the program GO 3538 (PI: Edoardo Iani) and the wide-band images taken by GTO 1208 (PI: Chris Willott), both using the JWST/NIRCam instruments and targeting the Abell 370 cluster.
The \oiiio{4959,5007\AA} and \hb{} lines are covered in the F182M filter ($\lambda_\mathrm{rest}$=4522--5168 \AA), and the \ha{} line is covered in the F277W filter ($\lambda_\mathrm{rest}$=6363--8225 \AA). 
The line emission maps of \oiii{} and \ha{} were created by subtracting the line-bearing images by the continuum maps interpolated from the images taken in the neighboring filters. 
The spatial resolution of NIRCam at 1.8 \um{} is $\sim$ 0.06'' with pixel scale 0.03''/pixel. 
The pixel-wise surface brightness sensitivity of the \oiii{} map is $1 \sigma_\mathrm{SB}=3.0 \times 10^{-17} \,\mathrm{erg\,s^{-1}\,cm^{-2}\,arcsec^{-2}}$. 
A detailed description of the data reduction and continuum subtraction can be found in Appendices~\ref{sec:robustness} and \ref{sec:cont_sub}, respectively. 


The \lya{} nebula was observed with the Keck Cosmic Web Imager (KCWI), reported in \citet{li19}. 
The seeing on the date of the observation is about 1.5''. 
The astrometry of the \lya{} data is corrected by the alignment to the central position of the quasar (L1) and the foreground elliptical galaxy SDSS J023952.03-013549.8 \citep{sdssdr6} to the north of the system. 
The emission line map is constructed by pixel-wise continuum-subtraction and integration over a 25\AA\, bandwidth, according to the process described in \citet{li19} and Appendix~\ref{sec:cont_sub}.

\section{Results}
\label{sec:results}

In Fig.~\ref{fig:oiii_map}, we show the extended emission features that are unique to the F182M images and compare them to the stellar continuum.
We highlight the line emission in the continuum-subtracted image, which we identify as \oiii{} with \hb{} contribution $\leq 12.5\%$ (Appendix~\ref{sec:line}). 
Due to the small fraction and large uncertainty of the \hb{} flux, in addition to the partial coverage of the \ha{} map, we did not make the correction for \hb{} in the F182M continuum-subtracted image; thus we refer to the image as the \oiii{} map. 

\begin{figure*}
\centering
\includegraphics[width=0.66\textwidth]{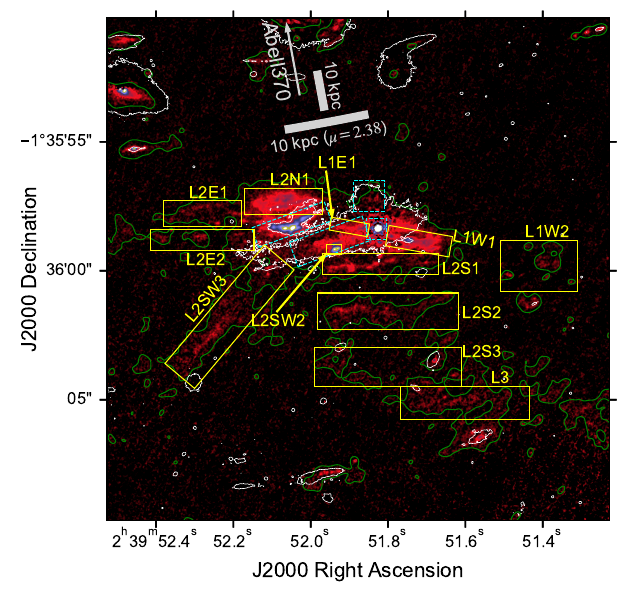}
\caption{Identification of the \oiii{} emitting features. The solid yellow boxes highlight the regions of the extended \oiii{} emission discussed in the work with labels. As a visual aid for identifying diffuse emission, we plot the solid green contours corresponding to the 3-$\sigma$ level ($1.8 \times 10^{-17} \,\mathrm{erg\,s^{-1}\,cm^{-2}\,arcsec^{-2}}$) of the \oiii{} line map when smoothed to a resolution of 0.3'', enclosing most of the reliable features of extended emission features. The solid white contours correspond to the stellar continuum in Fig.~\ref{fig:oiii_map} starting at 10-$\sigma$ flux density level $4.2 \times 10^{-2} \,\mathrm{MJy\,sr^{-1}}$. The dashed cyan boxes mark the same continuum sources as in Fig.~\ref{fig:oiii_map}. The grey arrow and scale bars are copied from Fig.~\ref{fig:oiii_map}. }
\label{fig:decomp}
\end{figure*}

\begin{figure*}
\centering
\includegraphics[width=0.66\textwidth]{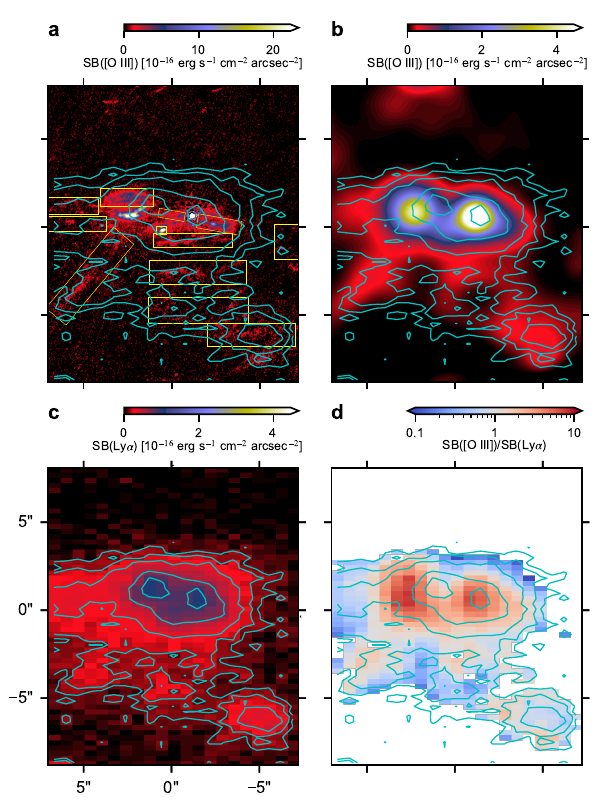}
\caption{Comparison between the \oiii{} and \lya{} line maps. {\textbf a}, \oiii{} surface brightness map overlaid with the \lya{} contours \citep[cyan;][]{li19}. The \lya{} contour levels correspond to +2, 3, 5, 10, and 15 $\sigma_\mathrm{SB}$ (Appendix~\ref{sec:cont_sub}). The yellow boxes mark the same line emitting features as shown in Fig.~\ref{fig:decomp}. {\textbf b}, \oiii{} map smoothed to the same resolution as the \lya{} image for comparison, overlaid with the \lya{} contours (cyan). {\textbf c}, \lya{} surface brightness map. {\textbf d}, \oiii{}-to-\lya{} surface brightness ratio map overlaid with the \lya{} contours (cyan). Only the pixels with the surface brightness larger than the 2 $\sigma_\mathrm{SB}$ for both \lya{} and the smoothed \oiii{} are plotted. }\label{fig:oiii_lya}
\end{figure*}

Thanks to the high spatial resolution and sensitivity, the [O III] line map resolves exquisite details of the CGM gas distribution. 
In contrast to the previous views of faint and diffuse CGM emission, the \oiii{} emission from the CGM appears bright and arise from several narrow filamentary structures. 
The decomposition and naming of the \oiii{} complex is shown in Fig.~\ref{fig:decomp} with detailed description in Appendix~\ref{sec:distribution}. 
In the vicinity of the system, there are two plumes (L1W1, L1E1) connected to the quasar, a bright compact source to the south (L2SW2), and clumpy gas clouds in the north of L2 (L2N1). 
The westwards plume (L1W1) extends for at least 8 kpc (after correction for lensing, same below) and could be linked to the tail at a projected distance of 25 kpc (L1W2). 
On large scales, fainter and filamentary features are found in several directions. 
East to L2, two narrow filaments extend for about 10 kpc (L2E1, L2E2). 
In the southeast, a bright filament (L2SW3) stretches from the DSFG (L2SW) with an extent of roughly 35 kpc. 
Directly to the south of the quasar (L1), three filaments (L2S1, L2S2, and L2S3\&L3) lie parallel to the DSFG-quasar system, with a projected separation of 15--20 kpc between each other. 
The patchy \oiii{} emission at a projected distance 60 kpc to the southwest (L3) coincides with a \lya{} emitting source \citep[][see also Fig.~\ref{fig:oiii_lya}]{li19}. 

The filaments display high \oiii{} surface brightness $1 - 2 \times 10^{-16} \,\mathrm{erg\,s^{-1}\,cm^{-2}\,arcsec^{-2}}$. 
They also appear to be similar in length (3--5'', 10--20 kpc) and width ($\sim$0.3'', 2.5 kpc). 
Following the parallel filaments (stripes) towards south, the surface brightness decreases and the appearance becomes more spread out and clumpy. 
It is also worth noting that the emission appears mainly in the southern half of the field, indicating asymmetric powering or matter distribution in the CGM. 

This newly discovered extended \oiii{} emission has both spatial scale ($\sim$ 100 kpc) and extension similar to those of the known \lya{} ``nebula'' (Fig.~\ref{fig:oiii_lya}), which extends towards the east and south and shows a tertiary peak at the southwest (L3) location. 
After smoothing the \oiii{} map to 1.5'' resolution to match with that of the \lya{} map, it can be seen that both \oiii{} and \lya{} display a nearly identical morphology on large scales. 
In addition to the morphology, the \oiii{} surface brightness also varies along with the \lya{} levels in the CGM. 
In the smoothed \oiii{}-to-\lya{} surface brightness ratio map in Fig.~\ref{fig:oiii_lya}, the regions in the outskirts of the host galaxies consistently show a ratio of unity, with variations of less than 0.3 dex. 
This ratio becomes much higher within the host galaxies (L1\&L2), and the \oiii{} peak is offset from the \lya{} peak in L2, due to the low fraction of \lya{} photons that can escape from the interstellar medium (ISM). 

\begin{figure*}
\centering
\includegraphics[width=\textwidth]{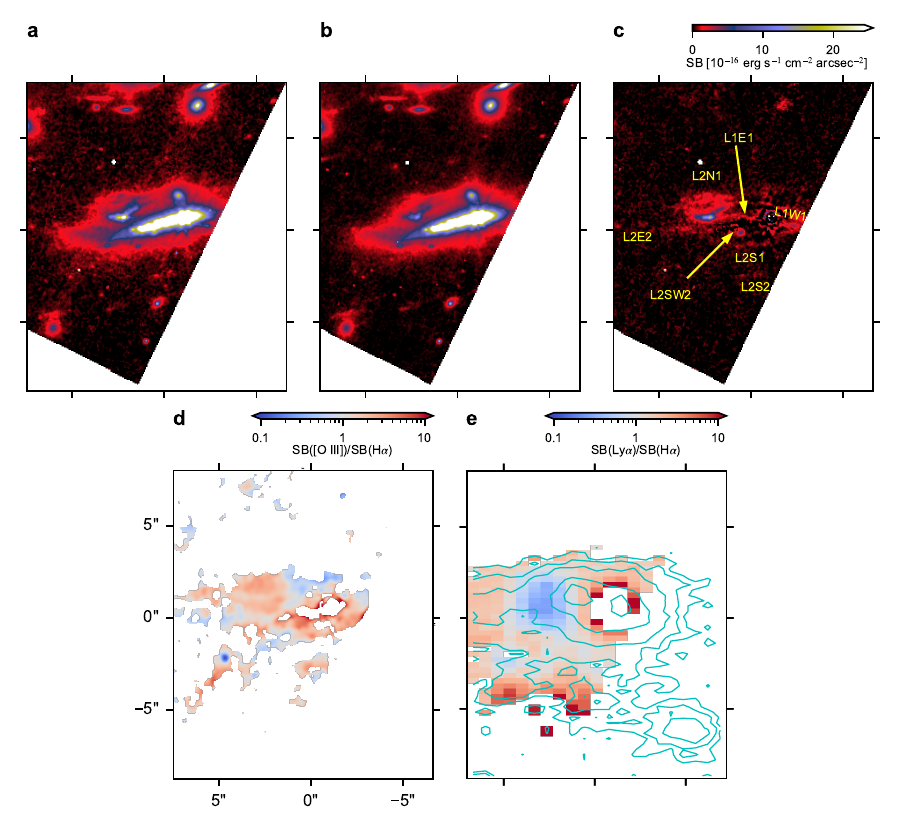}
\caption{Processing and analysis on the \ha{} line map. {\textbf a}, JWST F277W image. {\textbf b}, the interpolated stellar continuum using the F210M, F300M, F335M, and F356W images. {\textbf c}, \ha{} line emission map produced by subtracting {\textbf b} from {\textbf a}. Some features identified in \oiii{} (Fig.~\ref{fig:decomp}) that are also visible in \ha{} emission are labeled in the figure. {\textbf d}, \oiii{}-to-\ha{} surface brightness ratio map, after smoothing both lines to a common spatial resolution of 0.3''. {\textbf e}, \lya{}-to-\ha{} surface brightness ratio map after smoothing the \ha{} map to match with the resolution of the \lya{} map, overlaid with the \lya{} contours. For the two line ratio maps, only the pixels with surface brightness level higher than 2 $\sigma_\mathrm{SB}$ for both lines are used for making the line ratio map. }\label{fig:ha_map}
\end{figure*}

Some of the features identified in the \oiii{} map can also be seen in the \ha{} map (Fig.~\ref{fig:ha_map}) which has lower sensitivity and only partial coverage of the field. 
The surface brightness ratio between the \oiii{} and \ha{} images proves that the line emission of the extended regions in the F182M continuum-subtracted image is dominated by \oiii{}, with the potential contamination of \hb{} being less than $12.5\%$ (Appendix~\ref{sec:line}).

\section{Discussion}
\label{sec:discussion}

The high intrinsic surface brightness and the collisional excitation of the \oiii{} lines enable us to place unprecedented constraints on the physical conditions of the ``cool'' CGM gas. 
The \oiii{} line surface brightness is dependent on several parameters, including the gas column density, \Nh{}, the absolute metallicity, O/H, the ionization fraction of the emitting ion, O$^{2+}$/O, the electron density, \edens{}, and the electron temperature, \Te{}. 
However, a combination of the surface brightness level, the ratio of \oiii{}/\lya{}, and assumptions on the ionization fraction and geometry can significantly reduce the number of degrees of freedom, placing the constraints in Eq.~\ref{equ:constraints} below: 
\begin{subequations}
\begin{align}
   &\frac{\mathrm{SB}_\mathrm{[O\ III]}}{1.4\times10^{-16} \,\mathrm{erg\,s^{-1}\,cm^{-2}\,arcsec^{-2}}} \propto \left(\frac{n_e}{1 \,\mathrm{cm^{-3}}}\right)^2\left(\frac{20000 \,\mathrm{K}}{T_e}\right)^{1/2} \label{equ:constraints-oiii} \\
   &\mathrm{SB}_\mathrm{[O\ III]}/\mathrm{SB}_\mathrm{Ly\alpha} \propto \left(\frac{\mathrm{O/H}}{10^{-4.0}}\right)\left(\frac{\exp(-28700 \,\mathrm{K}/T_e)}{\exp(-28700 \,\mathrm{K}/20000 \,\mathrm{K})}\right) \label{equ:constraints-oiii_lya}
\end{align}
\label{equ:constraints}
\end{subequations}

The strong \oiii{} emission shows chemical enrichment at a large distance from the host galaxies. 
The joint constraints in Eq.~\ref{equ:constraints} is a strong function of \Te{} in the emitting gas, with an absolute lower limit for metallicity $\mathrm{O/H} > 10^{-4.6}$ (5\% \zsun{}) at high \Te{} end, while it increases to $10^{-3.4}$ (80 \% \zsun{}) if \Te{} is only $10^4 \,\mathrm{K}$. 
The almost equal brightness of \oiiio{4959,5007} and \lya{} emphasises the important role of metal lines in the cooling of CGM gas, especially in the surroundings of massive active galaxies or quasars that are enriched by feedback material. 
Despite the degeneracy, a reasonable solution to Eq.~\ref{equ:constraints} also requires high density up to $\sim 1 \,\mathrm{cm^{-3}}$ and pressure $\ge 2\times 10^4 \,\mathrm{K\,cm^{-3}}$, at least one order of magnitude larger than that of the cool CGM gas in typical cosmological simulations \citep[e.g.,][]{nelson20}. 
The individual dependence of the observed properties and the assumptions are detailed in Appendix~\ref{sec:condition}.

The \oiii{} and \ha{} line maps (Fig.~\ref{fig:ha_map}) help us solve the long-standing debate on the origin of \lya{} photons in the CGM--distinguishing between scattering and recombination. 
One direct indication is the small value of \lya{}-to-\ha{} surface brightness ratio (\lya{}/\ha{}), which reaches $\sim 3$ in the extended region and strongly disfavors the scattering scenario. 
This small value is consistent with the long-slit spectrum results on other \lya{} ``nebulae'' \citep{leibler18,langen23}, but the fact that it is smaller than the expected value of $\sim$ 8.7 implies either the limited sensitivity of the \ha{} map skews the ratio or a strong contamination of \nii{} and/or \sii{} in the same filter, or a fraction of \lya{} photons being scattering off the line of sight (LoS) due to the geometry, or the presence of considerable amount of dust obscuration in the CGM (see the detailed discussion in Appendix~\ref{sec:lya_ha}). 
On the other hand, the high \oiii{}/\lya{} and the high metallicity derived above indicate that most of the \lya{} photons are emitted in situ from the same ionized medium, as any contribution from the scattered \lya{} of the host galaxies would lead to a solution of an even higher metallicity by decreasing the intrinsic \lya{} surface brightness, $\mathrm{SB}_\mathrm{Ly\alpha}$, in Eq.~\ref{equ:constraints}; meanwhile, requiring a lower \Te{} where neutral hydrogen could exist. 
Therefore, we conclude that most of the \lya{} ``nebula'' emission are produced by recombination, and the lack of the scattered \lya{} from the host galaxies suggests the ``cool'' CGM gas around this system is dominant in the ionized phase. 

The extraordinary morphology and physical conditions further enable us to infer the processes that shape the CGM distribution and power the large-scale emission. 
Although photoionization is commonly hypothesised to be the powering source for the quasar nebulae \citep{cantalupo14,gonzalez23}, it is not favored in the context of this system because of the high \oiii{}/\ha{} line ratio, the geometry, and the small cross-section of gas filaments to reprocess the quasar radiation (see details in Appendix~\ref{sec:photoionization}). 
On the other hand, the high density and pressure require the gas to be either bounded by self gravity or pressurized externally by such processes as shocks or magnetic field. 
Evidence of shock heating in powering the CGM emission is seen in the irregular galaxy L2 and the surrounding gas structures, where the radio jet L1E1 likely collide with L2 and disrupt its ISM, resulting in bright line emissions enhanced by shocks as well as large clumps and tails of stripped gas (see the description in Appendix~\ref{sec:l2}). 
In the south, the parallel stripes suggest past episodic feedbacks separated by $\sim$ 20 Myr. 
However, the exact processes that shape and power these stripes are not clear, as scenarios including shock heating by the quasar jet, outflows driven by the starburst in DSFG, and the DSFG-quasar merger are all consistent in terms of the expected morphologies and timescales (calculations and comparisons in Appendix~\ref{sec:stripe}). 

We also note that the filaments surrounding SMM J02399-0136 resemble the \ha{} filaments found around local bright cluster galaxies (BCGs) that can also extend to beyond 50 kpc \citep[Appendix~\ref{sec:bcg};][]{conselice01,fabian03,olivares19}. 
These gas filaments play important roles in galaxy evolution as reservoirs of cool gas that may feed the ongoing or future star formation \citep{canning14,olivares19}. 
The morphology and the need for energy input also hints at the importance of large-scale magnetic fields \citep{fabian08,fabian11,fabian12,fabian16,tremblay18,olivares19,vigneron24}. 
However, SMM J02399-0136 CGM gas are in higher ionization states than typical BCG filaments \citep{tremblay18}, suggesting that it may represent a more energetic or earlier phase of the feedback. 

The inferred properties lead to a very different picture for this \lya{} ``nebula.'' 
The ``cool'' CGM in this system resides in high density and high pressure filaments with a low (0.3-0.5\% assuming $R_\mathrm{h}=150 \,\mathrm{kpc}$ typical for \lya{} nebula host halos) covering factor and extremely low ($\sim0.002\%$) filling factor, in contrast to the commonly accepted idea of widespread moderately dense clouds covering 10-50\% of the circumgalactic space \citep[c.f.][]{hennawi15,cai19}. 
The gas is also fully ionized through collisional ionization, at a temperature higher than $10^{4} \,\mathrm{K}$. 
Instead of tracing the accretion into the galaxy, the ``cool'' CGM gas emission in this system unveils past episodes of feedback. 

The importance of mechanical heating in shaping and powering CGM also poses challenges to numerical simulations, as the localized discontinuity created by shocks requires high spatial and time resolution in computation, in addition to the complex physical processes involved in outflows and jets \citep{bennett20,talbot24}. 
Even though shocks induced by feedback processes are common in cosmological simulations, the limitation in resolution and a lack of complete treatment of radiative cooling made previous studies mainly associate shocks with the ``hot'' ($10^{6} \,\mathrm{K}$) phase of CGM. 
Due to these limitations, simulations also tend to underestimate the density of the CGM while overestimating the cloud size and the photoionization rates in the feedback driven shocks. 
Contrary to the results of past simulations on cool accretion in the CGM \citep[e.g.,][]{rosdahl12}, this observation shows that dense clouds in the CGM could be dominantly ionized. 

The discovery of \oiii{} filaments show an unprecedented view of the ``cool'' CGM gas, but it is unclear whether this is common among the \lya{} ``nebulae'' or instead represents a rare example of strong recent feedback in SMM J02399-0136 system. 
Nevertheless, this study demonstrates the power of deep and high-resolution medium-band imaging with JWST in discovering and characterizing CGM emissions. 
A systematic survey targeting known \lya{} ``nebulae'' can offer useful knowledge about the prevalence of such feedback driven CGM structures; however, dedicated observations using high-resolution IFU instruments, such as JWST/NIRSpec IFU or ground-based IFU equipped with adaptive optics such as VLT/ERIS \citep{davis18}, are required to obtain a comprehensive understanding of the properties and kinematics of the ``cool'' CGM gas.

\section{Summary}
\label{sec:summary}

Surrounding the massive galaxy system SMM J02399-0136 at redshift 2.808, we serendipitously discovered large-scale \oiii{} emission arising from the CGM. 
Making use of the unprecedented resolution and depth of the JWST images, we have drawn the following conclusions.
\begin{itemize}
   \item The extended \oiii{} emission displays the same spatial extension as the \lya{} ``nebula.'' The intrinsic \oiii{} surface brightness reaches $1 - 2 \times 10^{-16} \,\mathrm{erg\,s^{-1}\,cm^{-2}\,arcsec^{-2}}$ and it is as important for cooling as the \lya{} line. 
   \item The ``cool'' CGM gas in this system resides in filaments. 
   \item Based on \oiii{} and \lya{} emissions, we infer the emitting gas is dense, enriched in metal, fully ionized, and experiencing high pressure in the CGM. 
   \item Overall, feedback processes shape and power the CGM gas distributions and emissions. 
   \item Future high-resolution IFU observations on a large sample can provide statistics and physical insights of the feedback-CGM interactions. 
\end{itemize}

\begin{acknowledgements}
We would like to thank the referee for the constructive comments.  
We also thank Guinevere Kauffmann, Eugene Churazov, and Edith Falgarone for the helpful discussions. 
This work is based on observations made with the NASA/ESA/CSA \textit{James Webb} Space Telescope. 
The data were obtained from the Mikulski Archive for Space Telescopes at the Space Telescope Science Institute, which is operated by the Association of Universities for Research in Astronomy, Inc., under NASA contract NAS 5-03127 for JWST. These observations are associated with the program \#1208 and \#3538. 
E.I. acknowledges funding from the Netherlands Research School for Astronomy (NOVA). 
C.F. is partially supported by NSF Grant \#1847892. 
G.S. is partially supported by NSF Grant AAG-1910107. 
\end{acknowledgements}

%
%

\bibliographystyle{aa}
\bibliography{sn-bibliography}

%

\begin{appendix} 

\section{JWST data reduction and the robustness of detections}
\label{sec:robustness}

We downloaded the raw data from the Mikulski Archive for Space Telescopes (MAST) portal, and processed them using the JWST pipeline version 1.14.0 and Calibration References Data System (CRDS) version 1223. 
Customized processing were implemented in different stages of the pipeline reduction, including snowball, wisp and stripe subtraction, following the procedures detailed in \citet{bagley23}. 
The Abell 370 catalog \citep{lagattuta19} is used for the astrometric alignment before combining all the dithered exposures. 

\begin{figure*}
\centering
\includegraphics[width=0.72\textwidth]{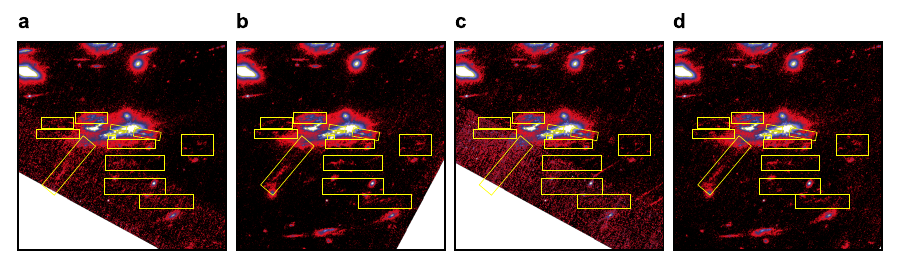}
\caption{Four individual F182M observations of the field, in order of observation association in MAST: o011 (\textbf{a}), o012 (\textbf{b}), o015 (\textbf{c}), and o016(\textbf{d}). The images were taken on the 19th (\textbf{a}), 22nd (\textbf{b}), 20th (\textbf{c}), and 19th (\textbf{d}) of December 2023. The images are shown in the same color scheme as Fig.~\ref{fig:oiii_map}}\label{fig:f182m_atlas}
\end{figure*}

Because the $1/f$ noise and the scattered light could strongly affect the low-surface-brightness features in the NIRCam image, we also show the four individual observations in Fig.~\ref{fig:f182m_atlas} for the F182M filter, with the data association codes o011 (a), o012 (b), o015 (c), o016 (d) in MAST, that have exposure time of 4638, 5068, 4638, and 5068 seconds, respectively. 
Although the galaxies are too close to the edge in the o011 and o015 exposures, and the southeast regions of the system show significantly higher noise due to the lower number of coverages in dithering for these observations associations. 
It can be seen that the faint filamentary features discussed in the paper are present in multiple exposures when the target is placed at different positions on the detector, confirming them as astrophysical features instead of ghost images due to scattered light or artefacts introduced in the data reduction.

\section{Continuum subtraction}\label{sec:cont_sub}

\begin{figure*}
\centering
\includegraphics[width=0.72\textwidth]{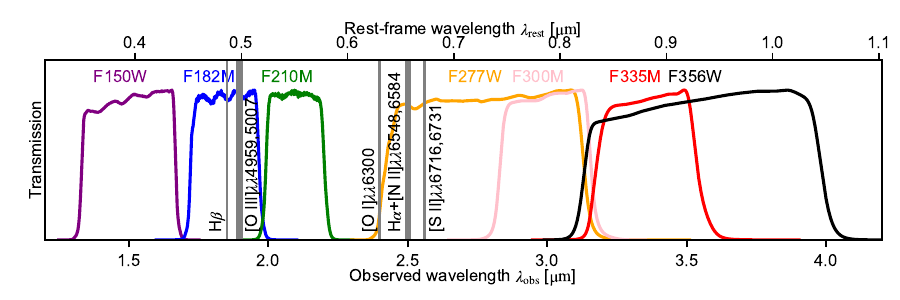}
\caption{Illustration of the filters used in the paper and their coverage of optical strong lines. The normalized transmission curves of each filters are plotted in solid lines, with the filter names labeled on top in the same color. The relevant optical strong lines (or complexes) are shown as vertical gray bands covering the corresponding wavelength ranges, with lines labeled next to the bands. }\label{fig:filters}
\end{figure*}

The stellar continuum is subtracted from the JWST images to show the distribution and the levels of the line emission. 
The continuum map for the F182M image ($\lambda_\mathrm{rest} \sim$ 4800 \AA) is interpolated as an average of the images in the neighboring bands F150W and F210M (Fig.~\ref{fig:filters}). 
Because the F150W image does not cover the southwest part of the field, the F210M image is scaled down by a factor of 0.85 to fill in this region. 
The scaling factor is estimated by comparing the interpolated continuum to the original F210M image in regions of common coverage. 
This rest-frame 4800 \AA\, continuum is subsequently subtracted from the F182M image, effectively removing most of the continuum emission (as shown in Fig.~\ref{fig:oiii_map}). 

A similar procedure is performed on the F277W image ($\lambda_\mathrm{rest} \sim$ 7300 \AA), while the continuum is interpolated as a weighted average of F210M, F300M, F335M, F356W images with weighting factors 0.5, 0.07, 0.07, and 0.36, respectively (Fig.~\ref{fig:filters}). 
The inclusion of several images accounts for the fact that although the F300M and F335M images are close to the F277W image in wavelength, the two exposures are short compared with the F210M and F356W images, leading to much noisier continuum image and are thus weighted down. 
The F210M data has higher resolution than the rest of the images used, partly because it is taken on the short wavelength detector, and partly because of the more dithered exposures than the other images. 
In addition, the pixel size of the short wavelength detection is only half of that of the long wavelength detector. 
Therefore the F210M image is smoothed and resampled to match with the resolution and the pixel size of the rest of the data when constructing the rest-frame 7300 \AA\, continuum. 

For the \lya{} cube, the continuum is modeled by averaging the spectral cube between the wavelength range $\lambda_\mathrm{obs}$ = 4450--5000 \AA, excluding 4515--4826 \AA. 
The continuum image is then subtracted from the \lya{} spectral cube, and the channels between $\lambda_\mathrm{obs}$ = 4615--4640 \AA\, ($\lambda_\mathrm{rest}$=1211.9--1218.5 \AA) are collapsed to create the \lya{} map. 

The pixel-wise surface brightness sensitivity, $\sigma_\mathrm{SB}$, is measured to be $3.0 \times 10^{-17}$ and $1.8 \times 10^{-17} \,\mathrm{erg\,s^{-1}\,cm^{-2}\,arcsec^{-2}}$ for the F182M and F277W continuum-subtracted images. 
We note that the sensitivity degraded rapidly near the edge of the F277W image due to a lack of coverage in dithered exposures. 
For the \lya{} cube, we use the sensitivity measured in \citet{li19}, corresponding to $\sigma_\mathrm{SB}$ = $6 \times 10^{-18} \,\mathrm{erg\,s^{-1}\,cm^{-2}\,arcsec^{-2}}$ measured with a 1.5'' aperture.

\section{Line identification}\label{sec:line}

The F182M filter covers the wavelengths of two strong optical line complexes \oiii{} doublet and \hb{} (Fig.~\ref{fig:filters}).
This degeneracy is solved by the comparison with the F277W data. 
As shown in Fig.~\ref{fig:ha_map}, the surface brightness of the line emission in the F182M continuum-subtracted image (\oiii{}\&\hb{}) is about three times that of the \ha{} in the F277W continuum-subtracted image in the extended regions discussed in this work. 
As the \ha{}/\hb{} flux ratio is expected to be $\sim$ 2.87 for the case-B recombination at $T_e \sim 10^4 \,\mathrm{K}$ \citep{hummer87}, we can constrain the contribution from \hb{} in the F182M continuum-subtracted image to be $\leq 12.5\%$, and conclude that the dominant line emission is from the \oiii{} doublet. 

The F277W image could be similarly contaminated by the neighboring \niio{6548,6584\AA}, and \siio{6717,6731\AA} lines (Fig.~\ref{fig:filters}), with \nii{} doublets being often the brightest lines. 
However, the \nii{}/\ha{} flux ratio is typically less than unity in star forming galaxies, reaching beyond one only in the high metallicity gas in the vicinity of AGNs \cite{baldwin81,kewley06}. 
Given that the CGM abundance is estimated to be order of magnitude lower than the AGN host galaxies ($\geq$ solar metallicity), lines other than \ha{}, especially \nii{}, could only account for percent levels of the line emission in the F277W continuum-subtracted image.

\section{[O III] distribution and morphology}\label{sec:distribution} 

\begin{table*}[]
\caption{Extended \oiii{} emitting features.}\label{tab:decomp}
\centering
\begin{tabular}{lcccc}
\hline
Name  & Right Ascension    & Declination  & Position Angle     & Integrated Line Flux\tablefootmark{a} \\
      & h:m:s (J2000)      & d:m:s (J2000)   & degree       & $10^{-16} \,\mathrm{erg\,s^{-1}\,cm^{-2}}$\\
\hline
L1W1  & 2:39:51.72      & -01:35:58.85 & -10          & 4.22$\pm$0.02\tablefootmark{b}  \\
L1W2  & 2:39:51.41      & -01:35:59.82 &              & 0.414$\pm$0.012 \\
L1E1  & 2:39:51.90      & -01:35:58.30 & -10          & 1.44$\pm$0.004\tablefootmark{b} \\
L2N1  & 2:39:52.07      & -01:35:57.31 &              & 5.97$\pm$0.008  \\
L2SW2 & 2:39:51.94      & -01:35:59.17 &              & 2.09$\pm$0.002  \\
L2E1  & 2:39:52.28      & -01:35:57.78 &              & 0.877$\pm$0.008 \\
L2E2  & 2:39:52.28      & -01:35:58.80 &              & 0.755$\pm$0.008 \\
L2SW3 & 2:39:52.21      & -01:36:01.78 & 50           & 2.58$\pm$0.014  \\
L2S1  & 2:39:51.82      & -01:35:59.74 &              & 1.97$\pm$0.029  \\
L2S2  & 2:39:51.80      & -01:36:01.56 &              & 1.59$\pm$0.013  \\
L2S3  & 2:39:51.80      & -01:36:03.73 &              & 0.873$\pm$0.014\tablefootmark{c}   \\
L3       & 2:39:51.60      & -01:36:05.13 &              & 1.52$\pm$0.012  \\
\hline
\end{tabular}
\tablefoot{Coordinates and flux of the extended structures in the continuum-subtracted F182M image (\oiii{} map) discussed in this work. The fluxes are integrated in the region marked by the yellow boxes in Fig.~\ref{fig:decomp}. \\
\tablefoottext{a}{Total flux of all the emission lines in F182M image; not corrected for gravitational lensing.}
\tablefoottext{b}{Partial confusion with quasar emission.}
\tablefoottext{c}{Excluding the line emitting galaxies in the same aperture.}}
\end{table*}

Fig.~\ref{fig:decomp} illustrates the features in and around SMM J02399-0136 as traced by the \oiii{} emission. 
The position and the \oiii{} fluxes of the extended regions are summarized in Table~\ref{tab:decomp}. 
The fluxes are integrated over the yellow boxes shown in Fig.~\ref{fig:decomp}. 
We adopt a flat $\mathrm{\Lambda CDM}$ cosmology with $\Omega_{\mathrm{m}}=0.28$ and $H_{0}=69.7 \mathrm{~km}\mathrm{~s}^{-1} \mathrm{~Mpc}^{-1}$ \citep{wmap}. 

The strong gravitational lensing by the foreground cluster Abell 370 distorts the appearance of SMM J02399-0136. 
However, at a distance of $\sim80^{\circ}$ from the potential centre (arrow in Fig.~\ref{fig:f182m_atlas}), the lensing acts by stretching the original image along the shear (orthogonal to the direction towards the lensing potential) by a factor of the magnification $\mu=2.38$. 
However, the length scale along the direction to the lensing potential and the intensity of the image remain roughly unchanged. 
Therefore, to infer the intrinsic shapes and physical scales of the system, one can simply imagine compressing the observed image by the magnification factor along the lensing shear, or use a polar-coordinate system with different scales along and orthogonal to the shear direction (scale bars in Fig.~\ref{fig:f182m_atlas}). 
For consistency, we report all the apparent sizes in observation as ``angular sizes'' (in arcsec), while referring to the (projected) physical lengths after lensing correction as ``distances'' or ``physical sizes'' (in kpc). 

The quasar (L1) appears as a point source in the \oiii{} map and dominates the flux in the quasar-DSFG merger system, while the \oiii{} emission from the DSFG (L2SW) is weak except for a few small clumps, most likely due to dust attenuation. 
At the south of the DSFG at a projected distance of $\sim$ 0.25'' (2 kpc), there is a small but bright elliptical source that does not have obvious continuum counterpart (L2SW2). 
The companion galaxy to the northeast (L2) displays a complex morphology in \oiii{}. 
The galaxy brightens significantly, showing two clumps in the central region. 
A prominent narrow extension towards northwest also exhibits strong line emission with only very weak continuum emission. 
Large amount of \oiii{} emitting gas lies at the north of the galaxy (L2N1) with high surface brightness and clumpy morphology. 
These bright clumps of gas extend for about 4 kpc across, comparable to the size of the galaxy L2 itself. 

Near the quasar, a collimated plume of gas (L1W1) appears to stem from the quasar nucleus towards the west direction. 
The plume shows a bright knot at a distance of 1.3'' ($\sim$ 5 kpc after correction for lensing) from the quasar, which turns into a more extended geometry at beyond 2'' ($\sim$ 8 kpc). 
Some \oiii{} gas clouds also exist in the same direction at a distance of 6--7'' ($\sim$ 25--30 kpc) (L1W2). 
These distant clouds resemble the tail of the same gas plume, based on their alignment and the morphology. 
In the opposite direction to the quasar, another stripe of \oiii{} emission (L1E1) extends from quasar towards L2 and L2N1. 

More \oiii{} emitting gas exists to the east of L2, forming two east-west filaments L2E1 and L2E2. 
L2E1 shows an arc-like geometry and seems to connect with L2N1. 
L2E2 comprises mainly of three clumps separated by 1.5'' (5 kpc). 
We discuss the possible jet-ISM origin of these features in the west in Appendix~\ref{sec:l2}.

A prominent \oiii{} structure (L2SW3) extends from the tidal tail of the DSFG towards southeast. 
The maximal extent of L2SW3 reaches a projected distance of 6'' (35 kpc) from the DSFG. 
L2SW3 shows mainly a bright filament for a length $\sim$ 4'' (24 kpc) and width 0.3'' (2.5 kpc). 
The far end of L2SW3 coincides with an elliptical galaxy lying at a lower redshift. 

Directly to the south of the DSFG-quasar merger, several bright filaments (stripes) can be seen. 
L2S1 lies at 1.3'' (10 kpc) south of the quasar, parallel to the DSFG-quasar system. 
It has a length of 4'' ($\sim$ 15 kpc) and width less than 3'' (2.5 kpc). 
The filament displays complex structures, including a few clumps and discontinuities. 
The west end of L2S1 appears to connect with L1W1 at an acute angle. 
L2S2 lies 2'' (16 kpc) south of L2S1, showing very similar geometry but at lower surface brightness. 
Its extension towards west appears to be in association with L1W2. 
Further 2.5'' (20 kpc) to the south, a filament extends in the northeast-southwest direction (L2S3 and L3). 
This structure is divided into two parts based on the \lya{} observation, as L3 coincide with a peak in the \lya{} map \citep{li19}, while L2S3 region traces a distinct component in the spatial-velocity space connecting L2SW3 to L2 \citep{vidal21}. 
Though L2S3 appears much dimmer than other filaments, the L3 has comparable surface brightness and geometry as L2S2. 

\begin{figure}
   \centering
   \includegraphics[width=0.48\textwidth]{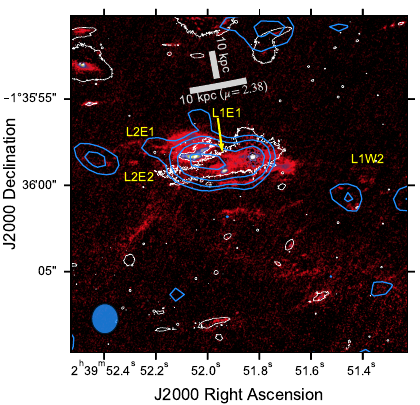}
   \caption{Map of radio emission. 1.4 GHz radio continuum is shown as blue contours, overlaid on the \oiii{} map in the background. The white contours and the grey scale bars represent the stellar continuum and 10 kpc size in the rest-frame, in the same format as Fig.~\ref{fig:decomp}. The names of some relevant \oiii{} emitting features are labeled as yellow text. The radio map contours correspond to +2, 4, 8, 16 $\sigma$, and the synthesized beam is shown at the lower left corner as a filled blue ellipse.}
   \label{fig:radio}
\end{figure}

A previous radio survey at 1.4 GHz identifies an extended (4'' $\times$ 1.1'') radio source between the quasar (L1) and the irregular galaxy (L2) \citep{ivison10,wold12}. 
For comparison, we plot the radio continuum over \oiii{} map in Fig.~\ref{fig:radio}, and labeled the relevant \oiii{} features in the figure. 
The description of the data can be found in \citet{ivison10}. 
At the center, the morphology of the radio emission appears elongated in the L1-L2 direction, enclosing the whole merging system. 
The direction of L1E1 and its spatial position embedded in the radio detection \citep{ivison10,wold12} suggest it may be the gas plume opposite to L1W1 launched by the radio jet. 
The lower \oiii{} surface brightness of L1E1 than L1W1 may be caused by the higher dust attenuation associated with the DSFG in this direction. 
In the east and west directions, two extended ``lobes'' coincide roughly with the locations of L1W2 and the end of the L2E1\&L2E2 tails.
The distant radio ``lobes'' suggest the potential role of past radio jets in creating these features. 

In addition to the structures surrounding SMM J02399-0136, more line emitting galaxies can be seen in the F182M continuum-subtracted image. 
Some of them display clear elongated shape similar to that of SMM J02399-0136, such as the smaller galaxies in the northwest of L2E1, the one directly south to L1W2, and the one south to L3. 
The galaxy to the northwest of L2E1 also shows up in the F277W continuum-subtracted image, assuring its association with SMM J02399-0136. 
However, a redshift confirmation of these galaxies requires dedicated spectroscopic observations. 
In addition to the small galaxies, some large ($> 1''$) galaxies also show traces of residual emission after the subtraction of their stellar continuum. 
Judging from their sizes, brightness of the continuums and lines, as well as the mostly elliptical morphologies, it is highly unlikely that they are field galaxies at comparable redshifts. 
We speculate that the residual line emission may correspond to the Paschen $\alpha$ line for the local galaxies, or the Paschen $\beta$ line for the galaxies in the Abell 370 cluster.

\section{Physical conditions in the CGM}\label{sec:condition}

The \oiii{} surface brightness is dependent on five physical properties \citep{osterbrock06,draine03,pyneb}, namely, gas column density, $N_\mathrm{H}$, metallicity or oxygen abundance, O/H, ionization fraction of the doubly ionized oxygen ion, ($\mathrm{O^{2+}/O}$), electron density, \edens{}, and electron temperature, \Te{}, as
\begin{equation}\begin{split}
   &\mathrm{SB}_{\mathrm{[O\,III]\lambda\lambda4959,5007}} \sim \\
   & 1.4 \times 10^{-16} \times \left(\frac{N_\mathrm{H}}{10^{22} \,\mathrm{cm^{-2}}}\right) \cdot \left(\frac{n_e}{1 \,\mathrm{cm^{-3}}}\right) \cdot \left(\frac{\mathrm{O/H}}{10^{-4.0}}\right) \cdot \left(\frac{\mathrm{O^{2+}/O}}{1.0}\right)  \\
   & \cdot \left(\frac{e^{-28700 \,\mathrm{K}/T_e}}{e^{-28700 \,\mathrm{K}/20000 \,\mathrm{K}}}\right) \cdot \left(\frac{20000 \,\mathrm{K}}{T_e}\right)^{1/2} \,\mathrm{erg\,s^{-1}\,cm^{-2}\,arcsec^{-2}}
\label{equ:i_oiii}
\end{split}.\end{equation}

Assuming the \lya{} photons as in-situ emission from the recombination of the ionized gas, the \oiii{}-to-\lya{} line ratio can be approximated as a function of O/H, $\mathrm{O^{2+}/O}$, and \Te{}. 
Using case-B recombination coefficients \citep{hummer87,draine11} and PyNeb \citep{pyneb}, the line ratio can be approximated as
\begin{equation}\begin{split}
   \frac{\mathrm{SB}_\mathrm{[O\,III]}}{\mathrm{SB}_\mathrm{Ly\alpha}} \sim 1.0 \times \left(\frac{\mathrm{O/H}}{10^{-4.0}}\right) \cdot \left(\frac{\mathrm{O^{2+}/O}}{1.0}\right) \cdot \left(\frac{e^{-28700 \,\mathrm{K}/T_e}}{e^{-28700 \,\mathrm{K}/20000 \,\mathrm{K}}}\right)
\label{equ:oiii-lya}
\end{split}.\end{equation}

The choice of the reference values in the formula are partly based on the consistency with the conditions inferred from other arguments; for example, beyond 20,000 K, collisional ionization starts to become relevant in producing O$^{2+}$ ions \citep{strawn23}.
This is partly based on the typical assumptions of CGM gas, such as an O/H value lower than the solar value \citep{lau16}. 
However, we present this scaling relation for the ease in numerical calculation and comparison, instead of showing a final solution. 

To further constrain the physical conditions in the emitting gas, we can make some heuristic assumptions. 
The first assumption is that the ionization fraction of the doubly ionized oxygen is unity:
\begin{equation}\begin{split}
   \mathrm{O^{2+}/O} = 1.0
\label{equ:o++-o}
\end{split}.\end{equation}
This is hardly the case in collisionally ionized gas \citep{mazzotta98}, and deviation from unity is more obvious when non-equilibrium collisional ionization is considered \citep{oppenheimer13}.
This naive assumption is a compromise due to the limited amount of information available, and future spectroscopic observations could enable us to better estimate $\mathrm{O^{2+}/O}$. 
Because of this assumption, the constraints on O/H and \Te{} in Eq. E.2 are actually lower limits. 
The second assumption concerns the morphology of the filamentary structure of the emitting CGM gas. 
All these filaments share a common narrow and elongated shape, despite different orientations relative to the quasar or the DSFG. 
Therefore, we make a simple assumption their intrinsic shape are also cylindrical with a width of about 3 kpc, which is far more likely than all of them being sheets viewed edge-on. 
This assumption enables us to relate the column density to the electron density, so that the width, $W$, corresponds to
\begin{equation}\begin{split}
   W \sim 3 \times \left(\frac{N_\mathrm{H}}{10^{22} \,\mathrm{cm^{-2}}}\right) \cdot \left(\frac{1 \,\mathrm{cm^{-3}}}{n_e}\right) \,\mathrm{kpc}
\label{equ:width}
\end{split}.\end{equation}

Combining the discussion above, we can place joint constraints on the physical conditions shown in Eq.~\ref{equ:constraints}.

\section{\lya{}/\ha{} and the effect of dust extinction}\label{sec:lya_ha}

Following the discussion of the small value of \lya{}/\ha{}, we explore the potential scenarios that could lead to the difference between the observed \lya{}/\ha{} and the theoretical value which varies only in the range 8.1--11.6 as a function of temperature and electron density \citep{hummer87,dijkstra19,langen23}. 
Firstly, the Milky Way foreground dust attenuation is expected to have negligible effect at the observed sky position at near-infrared wavelengths. 
But the observation sensitivity could severely bias the line ratio map to a lower value, as the signal-to-noise ratio threshold sets a ceiling for the 1/SB(\ha{}) term, and the value changes across the map due to degradation of sensitivity near the edge of the field. 
For example, the ratio of the threshold cuts $2\sigma_\mathrm{Ly\alpha}/2\sigma_\mathrm{H\alpha}$ is approximately 8 after being smoothed to the same spatial resolution, and the value further decreases in the edge pixels. 
Thus the value of \lya{}/\ha{} bears large uncertainty, and the interpretation of this value should be taken with caution. 
In addition, \nii{} and \sii{} lines fall in the same filter as \ha{} (Fig.~\ref{fig:filters}), and they could reach a similar level and even beyond in solar to super-solar metallicity strongly shocked medium \citep[c.f.][]{dopita05,allen08}. 
The strong low-ionization metal lines poses an explanation that is aligned with multiple evidences pointing towards mechanical heating, yet it is also non-trivial to coordinate with the metallicity estimated in the CGM gas. 

Despite the low fidelity and possible line confusion, the observed \lya{}/\ha{} $\leq$ 3 could also result from two astrophysical effects, which are the scattering of \lya{} photon off the LoS and the dust extinction. 
In the first scenario, considering that the emitting gas in filaments is optically thick for \lya{}, cylindrical filaments with elliptical cross-section, such that the major- and minor-axis are along and orthogonal to the LoS respectively with a ratio of 5:1, would be able to reproduce the fraction of \lya{} lost to scattering. 
This would increase our simple estimation for the gas column density in Eq.~\ref{equ:width}, while changing other parameters accordingly. 

If the loss of \lya{} is solely attributed to the dust extinction, at least 60\% of the \lya{} would be absorbed by dust based on the theoretical lower limit \lya{}/\ha{} $\ge$ 8.1. 
On one hand, it strengthens the arguments on the recombination dominated \lya{} production and the feedback enrichment in the CGM, and lifting the gas heating budget for the powering mechanism. 
However, this additional degree of freedom adds to the complexity of the quantitative characterization of the conditions in the CGM, especially the ones based on \oiii{}/\lya{}. 
Assuming that the intrinsic \lya{} is a factor of 3 higher, the parametric constraints in Eqs.~\ref{equ:constraints-oiii_lya} and \ref{equ:oiii-lya} would be relaxed accordingly, for instance, as $\log$ O/H = -4.5 (6\% \zsun{}), if the same $\mathrm{O^{2+}/O}$ and \Te{} are used. 
Because \oiii{}/\lya{} = 1 is adopted in Eq.~\ref{equ:oiii-lya} to derive Eq.~\ref{equ:constraints-oiii}, the latter would need to increase by a factor of 3 on the right-hand side. 
Therefore, the metallicity estimation for the system can be lower, but even higher pressure is required. 

Taking this assumed extinction of \lya{} and our simple estimation of ionized gas column density, we can estimate the dust column density as well as an upper-limit for the dust-to-gas ratio. 
Again, comparing \lya{}/\ha{} $\leq$ 3 to the lower limit of 8.1, we can derive $A_\mathrm{Ly\alpha}-A_\mathrm{H\alpha} = 1.08$. 
The type of dust surviving in the outflow to CGM is highly uncertain and is subject to active research \citep[e.g.][]{farber22,richie24}, so we use the extinction curve derived in \citet{calzetti00} ($R_\mathrm{V}$ = 4.05) that is widely used for DSFGs. 
This translates to $A_\mathrm{V}$ = 0.12 mag, though it is worth noting that because different extinction curves diverge significantly in the far-ultraviolet wavelength range, this value can change by a factor of 3 if a different type of dust is adopted. 
This extinction value is significantly higher than the typical excess of $A_\mathrm{V} \sim$ 0.003--0.005 mag measured for $z$=0.05--0.3 galaxies at impact parameter $\sim$ 60 kpc \citep{menard10,peek15}. 
Whereas the division to the estimated ionized gas column density gives a dust-to-gas ratio (DTG) $A_\mathrm{V}/N(\mathrm{H}) \sim 1.2 \times 10^{-23} \,\mathrm{mag\,cm^{2}}$, which is 50 times lower than those of the Milky Way and Large Magellanic Cloud, and 20 times lower than the Small Magellanic Cloud DTG \citep{predehl95,martin98,draine03}. 
Using the metallicity estimation of 6\% \zsun{} after the correction for dust extinction, the dust to metal ratio is estimated to be about 35\% that of the Milky Way, albeit comparable to the measurements of the high-z damped \lya{} absorbers \citep{decia13}. 
The small DTG is also aligned with the theoretical calculation that DTG can decrease by one order of magnitude in hot winds \citep{richie24}. 
Because only the ionized column is estimated and used in the calculation, we remind the reader that the total gas column density could be underestimated, making the derived DTG an upper-limit. 

Due to the limited information in imaging data, it is up to the future spectroscopic observations to distinguish between different scenarios.

\section{Inconsistency with photoionization}\label{sec:photoionization}

Even though it is strongly argued (in the previous section) that the extended \oiii{}, \ha{} and \lya{} emission arise from ionized gas, the exact mechanism that powers these emissions is not clear. 
Photoionization from the quasar has been widely conjectured as the powering source for CGM emissions, especially for the large \lya{} ``nebulae'' around quasars \citep{cantalupo14,gonzalez23}. 
However, multiple features observed in this system unfavor the role of quasar photoionization. 

For the CGM gas in the south, we can reject the quasar photoionization as a major powering source based on a simple argument of energy conservation. 
For distant filaments like L3, the observed high \oiii{} surface brightness $1.5 \times 10^{-16} \,\mathrm{erg\,s^{-1}\,cm^{-2}\,arcsec^{-2}}$ translates into a luminosity $5.2 \times 10^{41} \,\mathrm{erg\,s^{-1}}$ in every 1 kpc length segment unit, after correcting for gravitational lensing. 
The actual luminosity of the filament is at least twice the value above when adding up all the line emissions, e.g. \lya{} which is about as bright as [O III].
Therefore, the total energy in the QSO ionizing radiation received by the filament segment must at least exceed the combined luminosity of \oiii{} and \lya{}. 

The incident photoionization power on L3 can be estimated in different ways. 
Using the magnification corrected 5000\AA\, continuum, we can estimate the luminosity of the quasar to be $L_\mathrm{5000} \sim 5.0 \times 10^{44} \,\mathrm{erg\,s^{-1}}$, modulo the unknown dust attenuation effects. 
For our estimation, we assume an extreme case such that the central quasar ionizing luminosity that enters the CGM ($h\nu > 13.6 \,\mathrm{eV}$) is $5.0 \times 10^{44} \,\mathrm{erg\,s^{-1}}$ (corresponding to $Q_0 \sim 10^{54.8} \,\mathrm{s^{-1}}$), which is an extreme combination of the quasar luminosity, spectral energy distribution, and the ionizing photon escape fraction for a broad absorption line quasar like SMM J02399-0136, or that the CGM is illuminated by photons escaping through unobscured lines of sight in the quasar. 
Consider the furthest feature L3 at a projected distance of 60 kpc, which is a lower limit for the physical distance between the quasar and L3. 
Assuming a cylindrical geometry with width 3 kpc, the incident ionizing photon energy per 1 kpc length is $3.3 \times 10^{40} \,\mathrm{erg\,s^{-1}}$. 
 
Even if we consider a more extreme scenario that the QSO ionizing radiation in the L3 direction is completely unattenuated, which requires a highly anisotropic distribution of the dust and neutral gas, the distant filamentary features still lack enough illumination from the central QSO. 
In this case, we use the unobscured Lyman limit luminosity density $L_\mathrm{\nu, LL} = 1.2 \times 10^{30} \,\mathrm{erg\ Hz^{-1}}$ estimated in \citet{li19}. 
This corresponds to an ionizing luminosity about $L_\mathrm{ion} = 4 \times 10^{45} \,\mathrm{erg\ s^{-1}}$, and the energy received and re-emitted by each 1 kpc filament segment in L3 is $2.6 \times 10^{41} \,\mathrm{erg\,s^{-1}}$.  

Therefore, even with the unlikely assumption that QSO radiation on L3 is completely unobscured and that all the energy of the ionizing photon are converted to \lya{}+\oiii{} line emission, the quasar photoionization is only able to account for a small fraction of the observed bright CGM emission, falling short by at least 70\%. 
The photoionization scenario is fundamentally limited by the small filling factor of the gas clouds which makes them less effective in reprocessing the quasar radiation. (Appendix~\ref{sec:condition}). 
The problem is further worsened when considering the large amount of very hard photons (> 35.1 eV) and high ionization parameters needed to keep oxygen in the O$^{2+}$ forms. 

Another possible source for the photoionization of CGM gas is the cosmic UV background. 
However, the effect of the cosmic UV is negligible due to the high density of the emitting gas. 
Using the UV background $\Gamma_\mathrm{HI}=0.968 \times 10^{-12} \,\mathrm{s^{-1}}$ estimated at $z\sim2.8$ in \citet{faucher20}, the ionization parameter $U$ is $U \sim 5 \times 10^{-6} \left(1 \,\mathrm{cm^{-3}}/n_e\right)$. 
With $\log U \sim -5.2$, it is two orders of magnitude too low to ionize oxygen to be O$^{2+}$ in H II region-like conditions \citep{kewley01,morisset16}, not to mention that the hardness of the cosmic UV background is expected to be much lower than the stellar radiation that powers H II regions.

In addition, the observed value of \oiii{}/\ha{} is significantly higher than the predictions expected from the photoionization even for solar metallicity gas \citep{arrigoni15,kewley19}, while it could only be explained by the shock models at solar metallicity \citep{allen08}. 
It is even more challenging to invoke quasar illumination as the powering source for the extended emission to the southwest and to the west, since they are potentially blocked by the dust in the DSFG from the quasar sight lines. 
Therefore, in contrast to the previous studies on \lya{} ``nebulae'' that focused on quasar illumination, mechanical heating plays an non-negligible role in the CGM of SMM J02399-0136 system even at 60 kpc projected distance from the quasar.

\section{Heating in L2}\label{sec:l2}

L2 and the features surrounding it provide an example how the shock can directly shape and power the observed \oiii{} and \lya{} line emission. 
As described in Appendix~\ref{sec:distribution} and shown in Fig.~\ref{fig:radio}, the radio continuum display a large extension from L1 towards and beyond L2, potentially ending at the tail end of the eastward filaments L2E1\&L2E2. 
Therefore, we conjecture L1 launched a jet towards the direction of L2, whose direction is traced by both the radio emission and the narrow \oiii{} plume L1E1. 

Several lines of evidence, including the presence of the jet, the irregular shape of the companion galaxy L2, the brightened \lya{} and \oiii{} from the northwest extension of L2, the \oiii{} emission extending to the north (L2N1) and to the east (L2E1, L2E2), taken together support the view that the observations we present here are an aftermath of a relativistic jet clashing with and disrupting the ISM in L2.  
The northwest extension is the location of the direct interaction which shows enhanced \oiii{} brightness and peaks in the \lya{} map, while the clumps in the north (L2N1) are the ISM that is shock heated, blown away, and cooling through the \oiii{} line emission. 
The offset in the \lya{} and \oiii{} peak near L2 is most likely a result of the low escape fraction of \lya{} photons from the ISM in L2. 
The radio ``lobe'' and \oiii{} filaments further to the east (L2E1, L2E2) correspond to the past powerful jet and the gas it stripped away from the host galaxy L2, though the bulk of the gas outflow is lagged behind the relativistic jet. 

Although we caution that the star formation in the DSFG L2SW may have significant contribution to the radio continuum, whereas the resolution of the available data is not enough to disentangle different components. 
A detailed analysis of the relativistic jet in this system would require future VLBI observations.

\section{Heating, cooling, and the outflow rate in the parallel stripes.}\label{sec:stripe}

There are several mechanisms that may be responsible for the parallel filaments (stripes) seen in the south, including shock heating by the quasar jet, outflows driven by the starburst in DSFG, and the DSFG-quasar merger. 

Morphologically, the possible connections between the stripes (L2S1 \& L2S2) and the \oiii{} plume (L1W1 \& L1W2) could suggest that they are the shock fronts (cone) of the propagating jet. 
On the other hand, the parallel alignment between these stripes, the DSFG disk, as well as the DSFG-quasar merger, are also consistent with either the starburst or tidal interaction driven outflows. 

We can further estimate the timescale and outflow rate associated with the parallel stripes. 
We assume a typical speed 1000 km s$^{-1}$ for an \oiii{} wind, which is consistent with the linewidth of the \lya{} ``nebula,'' the linewidths of the ultraviolet (UV) emission lines of this quasar, as well as the velocity of a broad absorption line system in the (UV) spectrum of the quasar \citep{vernet01,li19}. 
Taking the projected distance 60 kpc, the travel time for the furthest point (L3) is about 60 Myr. 
Assuming a density of 1 cm$^{-3}$ and cylindrical geometry, each of the filamentary structures of size 2.5 kpc $\times$ 15 kpc contains $2.6\times10^9 \,\mathrm{M_\odot}$ ionized gas. 
Therefore, by considering only the three parallel stripes as the feedback remnants from the DSFG-quasar system, we can estimate an upper-limit for the outflow rate as 130 $\mathrm{M_\odot\, yr^{-1}}$. 

In the context of quasar feedback, the separation of the stripes represents the episodic nature of the AGN activity at an interval of about 15--30 Myr, and the ratio between the width and the separation can be naively interpreted as a duty cycle $f_\mathrm{duty}$ = 8--16 \%. 
These values are roughly consistent with estimations based on quasar number density and growth history \citep[e.g.,][]{martini04,eilers24}, though with large uncertainties. 
This timescale also agrees with the molecular cloud lifetime observed in nearby galaxies\citep{kruijssen19,chevance20} as well as the gas depletion time estimated for this galaxy \citep{vidal21}, and the estimated outflow rate is within a factor of 3 compared to the star formation rate. 
Thus, the parallel stripes can also be explained by either the scenario of episodic star formation such that the molecular clouds are destroyed by the stellar feedback and a certain period elapses before the gas is cooled down again for another round of star formation, or the episodic starburst scenario where the galaxy is alternating between fast accumulation of molecular gas and intense star formation that consumes the former and halts inflow from CGM. 

In the outflow scenario, the kinetic energy in the gas matches roughly as the radiative energy over the outflow timescale. 
Using the outflow rate estimated before, we can infer a kinetic energy of $2.6 \times 10^{58} \,\mathrm{erg}$ in each stripe when it was launched in the galaxy. 
On the other hand, cooling in each stripe through \oiii{} and \lya{} lines has at least a luminosity $L = 1.5 \times 10^{43} \,\mathrm{erg\,s^{-1}}$, and the corresponding radiative cooling timescale is less than 55 Myr, comparable to the estimated travel time. 
However, this simple calculation ignores the fraction of the initial kinetic energy that would be used in work against gravity to reach such distance, and it also assumes that the physical distance between the objects is the same as the projected distance on the sky. 
The estimation also bears a large uncertainty in the initial outflow velocity, as doubling the initial velocity would quadruple the kinetic energy while halving the expected travel time, making the radiative dissipation a smaller fraction of the total energy budget. 

Besides, the estimated orbital period for the DSFG-quasar merger also falls into the same range as the stripe separation (Appendix~\ref{sec:stripe}). 
The dynamical mass of the DSFG can be estimated from the CO($J$=3-2) observation \citep{frayer18}. 
Using a linewidth $\sigma_v=500 \,\mathrm{km\,s^{-1}}$ and observed CO disk radius 3 kpc ($\sim$ 1''), the dynamical mass of L2SW is estimated as $M_\mathrm{dyn} = 5.2 \times 10^{11} \,\mathrm{M_\odot}$, twice of the estimated molecular gas mass. 
Based on multi-band images and various gas tracers, the quasar does not account for a significant fraction of mass in the system. 
Thus taking a conservative estimation that dark matter mass is equal to the baryon within the sphere spanning from L2SW center to L1, we can derive a reduced mass for the merging system as $M_\mathrm{sys} \sim 10^{12} \,\mathrm{M_\odot}$. 
As a heuristic calculation, we use a semi-major axis $a$ = 3 kpc and zero eccentricity, then the orbital period for the DSFG-quasar merger can be estimated as
\begin{equation}\begin{split}
   T = 1.5 \times 10^{7} \times \left(\frac{a}{3 \,\mathrm{kpc}}\right)^{3/2} \cdot \left(\frac{10^{12} \,\mathrm{M_\odot}}{M_\mathrm{sys}}\right)^{1/2} \,\mathrm{yr}
\label{equ:period}
\end{split}.\end{equation}

Therefore, it is difficult to pin down the exact process responsible for the observed CGM gas distribution in this system, and the AGN activity and starburst may be fundamentally driven by the merging process. 
Future spectroscopic observations can help us learn the kinematics of these filaments, and test these scenarios with the linewidths as well as their connections to the host galaxies in the velocity space.

\section{Comparison with BCG filaments}\label{sec:bcg}

We note that similar filaments of ionized gas have been found through \ha{} emission surrounding local BCGs such as Perseus A, extending for 10 to 50 kpc \citep{conselice01,fabian03}. 
The origin of these filaments is still subject to active study, while the morphology and kinematics of the gas favor the scenario that they are compressed by the active galactic nuclear (AGN) feedback, though other heating mechanisms including magnetic fields and cosmic rays may also be important in powering these filaments \citep{fabian08,fabian11,fabian12,fabian16,tremblay18,olivares19,vigneron24}. 
In addition, low frequency radio observations have found synchrotron-emitting filaments at intracluster medium (ICM) scale ($\ge$ 200 kpc), as a result of the interaction between radio jets with the magnetic filaments in the ICM \citep{rudnick22}. 
Therefore, we suggest that high-resolution, low-frequency radio observations can help us better understand the nature of these filaments, by revealing the exact location of the quasar jets as well as the potential existence of radio filaments around this system. 
From a theoretical point of view, these cooling stripes are also reminiscent of recent numerical studies \citep[e.g.,][]{thompson16,gronke22,faucher23} picturing the cooling shocks and the rebirth of ``cool'' clouds as feedback gas mixes with hot CGM while decelerating at the same time. 

These filaments play important roles in the evolution of the host galaxy system. 
The estimated outflow rate already accounts for $\sim$ 1/3 of the star formation rate in the system. 
And given the previous studies that find the outflow mass is often dominated by molecular gas \citep{tombesi15,feruglio15}, the feedback revealed by the stripes is able to expel a significant fraction of gas in the host galaxies and quench the star formation. 
Around local BCGs, extended massive molecular filaments or clumps are found in close spatial correlation with the \ha{} filaments, hypothesized as gas clouds that are uplifted by feedback and will flow back onto the host galaxies, fuelling future star formation \citep{olivares19}. 
Besides, streaks of young stars or star clusters exist at locations near but offset from the BCG \ha{} filaments, which are likely results of in-situ star formation of the gas filaments, and could feed the growth the stellar halos of the host galaxies \citep{canning14}. 
Although \oiii{} filaments around SMM J02399-0136 are different from the BCG \ha{} filaments in several ways, including their higher ionization state, the larger extent with respect to the mass of the group or cluster system, as well as the absence of stellar continuum near the filaments; these \oiii{} filaments observed at redshift $z \sim$ 2.8 could be at an earlier and/or more energetic stage of the CGM gas filaments formation dominated by recent feedback events, while the local BCG \ha{} filaments represent an evolved and cooler phase that is able to refuel the AGN and star formation activities in and around the host galaxies \citep{tremblay18}.

\end{appendix}
\end{document}